\documentclass[twocolumn,showpacs,preprintnumbers,amsmath,amssymb,superscriptaddress]{revtex4}

\usepackage{graphicx}
\usepackage{epsfig}
\usepackage{epstopdf}
\usepackage{dcolumn}
\usepackage{bm}
\usepackage{amsmath}
\usepackage{xspace}
\usepackage{multirow}
\usepackage{natbib}
\usepackage{color}

\begin{document}

\preprint{}

\title{Spin-Momentum Locked Topological Surface States, non-trivial Berry's phase and magnetoelectric quantization in topological insulators}

\author{M. Zahid Hasan}
\affiliation{Joseph Henry Laboratories, Department of Physics, Princeton University, Princeton, NJ 08544, USA}

\date{\today}
ÊÊÊÊÊÊÊÊÊÊÊÊ



\begin{abstract}
The surface of a 3D topological insulator is a new type of two dimensional electron gas (2DEG). The exactly quantized magneto-optical or axionic response of a 3D topological insulator is a direct result of the Berry phase protected spin-momentum locked states on its surface. (Viewpoint on PRL {\bf 105}, 057401 (2010))


\end{abstract}

\maketitle
The excitation of macroscopic quantum matter often occurs in lumps: The amount of magnetic flux that pierces a superconductor can only increase in units of the flux quantum of $h/2e$; the conductance of a two-dimensional electron gas in a magnetic field is quantized in units of $e^2/h$ (the conductance quantum). Now, writing in {\it Physical Review Letters} \cite{Tse}, Wang-Kong Tse and Allan H. MacDonald of the University of Texas at Austin, US, present theoretical calculations that show that the magneto-optical response of a three-dimensional topological insulator - an otherwise nonconducting material with a band structure that gives rise to conducting states along its surface - is quantized in units of the vacuum fine-structure constant, $\alpha$=$e^2/\hbar c$=1/137. Their finding is an example of how the exotic properties of topological order in a three-dimensional solid lead to an exactly quantized excitation \cite{Hsieh, Hsieh1, Xia, Hsieh2, Roush, Nish, Hsieh3, Chen, Lin, Moore, Hasan,Qi}.

The most well-known example of a topological phase is a 2D electron gas at low temperatures and in a high-magnetic field, which has a quantized Hall conductance. However, the conducting states on the surface of a 3D topological insulator, while bearing some similarities to those in the 2D case, are a new state of matter  \cite{Moore, Hasan,Qi, Fu, Qi1,Essin, Day}. Theorists are therefore keen to describe the exotic physical properties of 3D topological insulators, which should exhibit new quantization rules, and predict the ways in which they can be observed in experiments.

The distinction between topological insulators and conventional band insulators is evident in the space that contains their allowed wave functions, i.e., the Hilbert space. In some sense, topological insulators are defined by the fact that their Hilbert space topology cannot be easily perturbed to destroy the wave functions on the surface. Hence the surface electron modes are protected.

\begin{figure}[htbp]
\begin{center}
\includegraphics[width=3in]{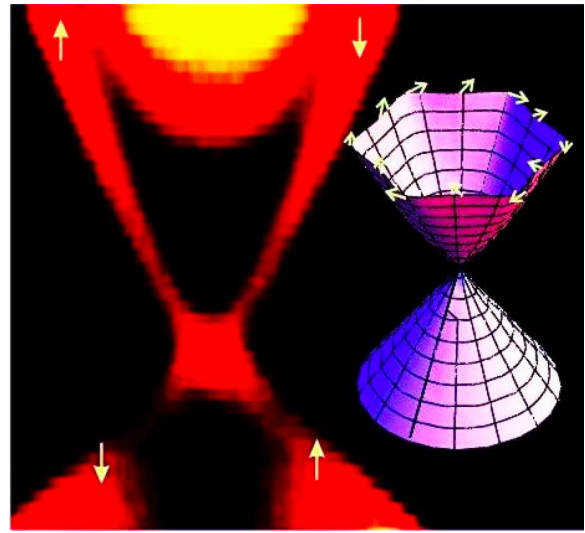}
\caption{\label{Figure1} \textbf{A new type of 2DEG}: The energy vs momentum of the states on the surface of a 3D topological insulator forms a spin-textured Dirac cone. The red curve is a slice through this cone, measured with photoemission from the surface of the topological insulator Bi$_2$Se$_3$. (Inset) Schematic of the full Dirac cone and the circulation of the spin states (yellow arrows) associated with each momentum state. The spin circulation pattern leads to a Berry's phase of $\pi$. (Illustration: Reproduced from Hsieh et.al., \cite{Hsieh2}).
}
\end{center}
\end{figure}

In the simplest description, the surface electron modes of a topological insulator are arranged in a single Dirac cone - the linear dispersion that describes massless particles - with a vortexlike spin arrangement (Fig.1). The circulating structure of the spins contributes a Berry's phase of $\pi$ to the electronic (or hole) wave function \cite{Hsieh1, Xia, Hsieh2, Roush} (recall that a spin-1/2 particle must undergo two complete rotations to acquire a phase of 2$\pi$). The Berry's phase protects these surface states against backscattering from disorder and impurities and dictates new topological quantization rules  \cite{Qi1, Essin}.

The spin vortex like pattern on the surface of a topological insulator exists in the presence of time reversal symmetry, but when this symmetry is broken - say, due to the presence of a magnetic field - a gap will open in the Dirac spectrum that disrupts the spin-texture. This can lead to unusual electromagnetic and magnetotransport effects in a topological insulator.

These effects can be quite spectacular when a magnetic field is applied perpendicular to the surface of a topological insulator. A magnetic field will induce Landau levels - the quantized states that give rise to the quantum Hall effect - to appear in the surface electronic spectrum. The Landau levels for Dirac electrons are special, however, because a Landau level is guaranteed to exist at exactly zero energy. Since the Hall conductivity $\sigma_{xy}$ increases by a conductivity quantum of $e^2/h$ when the Fermi energy crosses a Landau level, the presence of a Landau level at zero energy means the conductivity must be half-integer quantized: $\sigma_{xy}$=$(n+1/2)e^2/h$.

This behavior has been famously demonstrated in experiments on graphene, except that the Dirac points in graphene have a fourfold spin and valley degeneracy, which means the observed Hall conductivity is still integer quantized. At the surface of a bulk topological insulator, however, there is only a single spin-polarized Dirac cone (i.e., one that circulates in a particular direction) that carries ¹ Berry's phase (Fig. 1). This \textquotedblleft fractional\textquotedblright $\hspace{0.01cm}$ integer quantized Hall state [$\sigma_{xy}$=(1/2)$e^2/h$ for $n$=0] on the surface should be a cause for concern because the integer quantized Hall effect is always associated with chiral edge states (as opposed to helical surface states) and can only be integer quantized. The resolution is the mathematical fact that a surface cannot have a boundary. If the topological insulator is shaped like a slab (Fig. 2), the top surface and bottom surface are necessarily connected to each other, and will always be measured in parallel in transport, doubling the 1/2. The top and bottom can share a single chiral edge state, which carries the integer quantized Hall current.

A similar surface quantum Hall effect, known as the anomalous quantum Hall effect, can be induced with the proximity to a magnetic insulator. A magnetic field - say, from a nearby thin magnetic film - on the surface of a topological insulator lifts the spin degeneracy at the surface Dirac point. If the Fermi energy is in this induced energy gap, this leads to a half-integer quantized Hall conductivity $\sigma_{xy}$=$(1/2)e^2/h$ (Fig. 2) due to the Berry's phase of $\pi$ on the topological surface.

To induce such a gap, Tse and MacDonald propose to place a thick film of a topological insulator between two ferromagnets. They then consider the electromagnetic response of a helical (spin-momentum locked) Dirac gas in a half-integer quantized Hall state. They showed that under such conditions, the polarization of light transmitted through a 3D topological insulator would always be rotated by a fixed angle of tan$^{-1}(\alpha)$: rotation by any arbitrary angle is not possible. The flip side of this transmission or \textquotedblleft Faraday effect\textquotedblright $\hspace{0.01cm}$ is a quantized \textquotedblleft Kerr effect\textquotedblright , $\hspace{0.01cm}$ where light reflected from the surface has its polarization rotated by a fixed amount of tan$^{-1}(1/\alpha)$. The only requirement is that the light be at a frequency lower than the topological insulator band gap and the induced magnetic gap on the Dirac states.

\begin{figure}[htbp]
\begin{center}
\includegraphics[width=3in]{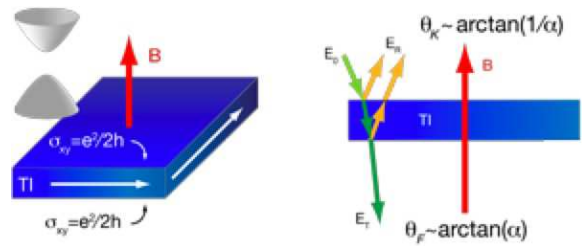}
\caption{\label{Figure 2} Magneto-optical effect in a 3D topological insulator (TI). (Top) Applying a weak magnetic field to the surfaces of a 3D topological insulator (blue slab) causes an energy gap to open in the spectrum of electronic surface states (grey). This amounts to a half-integer Hall conductivity ($\sigma_{xy}$) on each of its surfaces. (Bottom) The rotation of light that is reflected from (Kerr effect) or transmitted through (Faraday effect) a 3D topological insulator in a magnetic field is given by arctan(1/$\alpha$) and arctan($\alpha$), respectively. These angles are independent of material details. $E_0$, $E_R$, and $E_T$ represent incident, reflected, and transmitted beams. $\theta_F$ and $\theta_K$ are the Faraday and Kerr angles, respectively. (Illustration: Alan Stonebraker) }
\end{center}
\end{figure}

This unusual optical response of 3D topological insulators comes from a combination of cavity confinement and the Hall conductivity of spin-helical Dirac modes on the surface: The incident light can only induce quantized currents with a certain spin helicity, which in turn affects the reflected and transmitted light. The fine structure constant $\alpha$ dictates the coupling between the quantized currents and the electromagnetic wave. The Kerr rotation of tan$^{-1}(1/\alpha) {\sim} $tan$^{-1}$(137) $\sim{\pi}$/2 suggests that polarization of the reflected light should exhibit a striking full-quarter rotation relative to the incident polarization direction of the light beam (Fig. 2).

The effect Tse and MacDonald predict is effectively insensitive to the precise value of the gap as long as it is finite, since one can always choose a lower frequency of the incident light. Working with far infrared light, these conditions are adequately met in the topological insulator Bi$_2$Se$_3$ \cite{Xia, Hsieh2}. This highly tunable material features almost ideal Dirac quasiparticle helical spin modes that are locked in by the Berry's phase of $\pi$, as in Fig. 1. The surface modes are well protected within a large band gap ($\sim$ 0.3 eV). To measure the magneto-optical effects directly, a film of Bi$_2$Se$_3$ would have to be thick enough that the surface electrons do not tunnel between the top and bottom surfaces, but films of this thickness could, in principle, be grown using molecular beam epitaxy.

Tse and MacDonald's proposal for measuring topological quantization in units of the vacuum fine structure constant $\alpha=e^2/\hbar c$ $\hspace{0.01cm}$ could lead to a new metrological standard for fundamental physical constants \cite{Qi, Fu, Qi1,Essin, Day}. In addition, observing topological quantization in Bi$_2$Se$_3$ would be confirmation, independent of spin-resolved photoemission, that topological order can exist in \textquotedblleft ordinary\textquotedblright $\hspace{0.01cm}$ bulk solids. In the long run, classifying bulk solids in terms of topological quantization may turn out to be a more powerful method of identifying phases of matter beyond the standard Landau paradigm, which is based on the idea of spontaneously broken symmetry. Perhaps the most significant \textquotedblleft effect\textquotedblright $\hspace{0.01cm}$ is that which the discovery of topological insulator states has had on research itself: physicists can now study the interplay of \textquotedblleft topological order\textquotedblright $\hspace{0.01cm}$and \textquotedblleft broken-symmetry order\textquotedblright $\hspace{0.01cm}$ in real experiments and with real materials that are, in principle, accessible to anyone \cite{Hasan}.

\end{document}